\documentclass[aps,prx,floats,twocolumn,showpacs,superscriptaddress,nobibnotes,10pt,floatfix]{revtex4-2}
\usepackage[svgnames]{xcolor} 
\usepackage{graphicx,epsfig}
\usepackage{times} 
\usepackage{amssymb,amsmath,multirow,rotate,color}
\usepackage{bm}
\usepackage{xcolor}
\usepackage[normalem]{ulem}

\definecolor{albicocca}{rgb}{0.98, 0.7, 0.2}
\definecolor{internationalorange}{rgb}{1.0, 0.31, 0.0}
\definecolor{giocolor}{RGB}{0, 150, 100}

\usepackage{lipsum} 
\usepackage{ragged2e}
\usepackage{soul}
\DeclareMathAlphabet\mathbfcal{OMS}{cmsy}{b}{n}

\begin{document}


\title{Hyperedge Overlap drives Synchronizability of Systems with Higher-Order interactions}

\author{Santiago Lamata-Otín}
\thanks{These two authors contributed equally}
\affiliation{Department of Condensed Matter Physics, University of Zaragoza, 50009 Zaragoza, Spain}
\affiliation{GOTHAM lab, Institute of Biocomputation and Physics of
Complex Systems (BIFI), University of Zaragoza, 50018 Zaragoza, Spain}
\author{Federico Malizia}
\thanks{These two authors contributed equally}
\affiliation{Network Science Institute, Northeastern University London, London E1W 1LP, United Kingdom}

\author{Vito Latora}\email{v.latora@qmul.ac.uk}
\affiliation{Department of Physics and Astronomy,  University of Catania, 95125 Catania, Italy}
\affiliation{School of Mathematical Sciences, Queen Mary University of London, London E1 4NS, United Kingdom}
\affiliation{Complexity Science Hub Vienna, A-1080 Vienna, Austria}


\author{Mattia Frasca}\email{mattia.frasca@dieei.unict.it}
\affiliation{Department of Electrical, Electronics and Computer Science Engineering, University of Catania, 95125 Catania, Italy}

\author{Jes\'us G\'omez-Garde\~nes}\email{gardenes@unizar.es}
\affiliation{Department of Condensed Matter Physics, University of Zaragoza, 50009 Zaragoza, Spain}
\affiliation{GOTHAM lab, Institute of Biocomputation and Physics of
Complex Systems (BIFI), University of Zaragoza, 50018 Zaragoza, Spain}

\date{\today}


\begin{abstract}
The microscopic organization of dynamical systems coupled via higher-order interactions plays a pivotal role in understanding their collective behavior. In this paper, we introduce a framework for systematically investigating the impact of the interaction structure on dynamical processes. Specifically, we develop an hyperedge overlap matrix whose
elements characterize the two main aspects of the microscopic organization of higher-order interactions: the inter-order hyperedge overlap (non-diagonal matrix elements) and the intra-order hyperedge overlap (encapsulated in the diagonal elements). This way, the first set of terms quantifies the extent of superposition of nodes among hyperedges of different orders, while the second focuses on the number of nodes in common between hyperedges of the same order. Our findings indicate that large values of both types of hyperedge overlap hinder synchronization stability, and that the larger is the order of interactions involved, the more important is their role. Our findings also indicate that the two types of overlap have qualitatively distinct effects on the dynamics of coupled chaotic oscillators. In particular, large values of intra-order hyperedge overlap hamper synchronization by favouring the presence of disconnected sets of hyperedges, while large values of inter-order hyperedge overlap hinder synchronization by increasing the number of shared nodes between groups converging on different trajectories, without necessarily causing disconnected sets of hyperedges.
\end{abstract}

\maketitle


\section{Introduction}
\label{sec:I}

In recent decades, the study of the structural organization of complex systems and its influence on their collective behavior has raised increasing attention. This surge of interest has been driven by advances in complex network science, which aims to elucidate emergent phenomena arising from the microscopic interactions encapsulated in the form of a graph~\cite{newman2003structure,latora2017complex}. More recently, network science has focused on higher-order, or group, interactions, recognizing their substantial impact on the dynamics and emergent properties of interacting systems~\cite{battiston2020networks,majhi2022dynamics,bick2023higher}. Research in this area spans fields as diverse as ecology~\cite{grilli2017higher}, social contagion~\cite{iacopini2019simplicial,matamalas2020abrupt,de2020social}, game theory~\cite{alvarez2021,civilini2021dilemmas,civilini2023explosive} and synchronization~\cite{tanaka2011multistable,millan2020explosive,skardal2020higher,gambuzza2021stability,gallo2022synchronization}. 

The exploration of dynamical systems with higher-order interactions began with the development of basic models designed to capture the influence of groups on the emergent phenomena~\cite{iacopini2019simplicial,skardal2020higher}. In order to facilitate analytical derivations, the structures under consideration were either random hypergraphs, a natural extension of random graphs to higher-order interactions, or random simplicial complexes. The latter group structures introduce a downward closure, meaning that all interactions involving subgroups of nodes of an existing interaction must also be present in the structure~\cite{battiston2020networks}. 

After the initial models of hypergraphs were introduced, subsequent research evolved towards more elaborate representations of higher-order interactions.~\cite{malizia2023hyperedge,jhun2019simplicial,landry2020effect,st2022influential,zhang2023higher,kim2023contagion,burgio2023triadic,ren2023complementary,kim2024higher}. Early findings in this direction revealed that the interaction strength alone does not dictate behavioral changes in social systems~\cite{jhun2019simplicial,landry2020effect,st2022influential}. Instead, factors such as the group size distributions and the heterogeneity in node participation in groups were shown to be relevant. Thereafter, the effects of different degrees of relaxation of the downward closure in simplicial complexes were explored. Studies by Kim et al.~\cite{kim2023contagion} and Burgio et al.~\cite{burgio2023triadic} showed that the inclusion level among different-sized interactions influences the nature of the transition to active states in the context of social contagion. In addition, Zhang et al.~\cite{zhang2023higher} investigated the impact of higher-order interactions on synchronization of Kuramoto oscillators, finding that hypergraphs favour synchronization more effectively than simplicial complexes. Moreover, Malizia et al.~\cite{malizia2023hyperedge} examined the correlations among hyperedges of the same size, showing how the nature of the transition towards emergent collective states, such as synchronization or epidemic outbreaks, varies from first to second order depending on the overlap among hyperedges of the same size.

All the previously mentioned works were inspired to the structural features observed in real higher-order structures. Indeed, at a microscopic level, real higher-order systems exhibit considerable complexity in terms of overlap of hyperedges~\cite{malizia2023hyperedge,landry2024simpliciality,lee2021hyperedges,ha2023clustering}. Specifically, Lee et al.~\cite{lee2021hyperedges} quantified this property by introducing the \textit{overlapness}, a metric measuring the ratio between the number of nodes belonging to hyperedges with and without repetitions, showing that this ratio greatly varies between real-world datasets. Deepening in this heterogeneity, Landry et al.~\cite{landry2024simpliciality} employed a set of metrics to show that some real-world structures resemble simplicial complexes whereas others show an organization closer to that of random hypergraphs. Finally, Malizia et al.~\cite{malizia2023hyperedge} introduced a measure, namely the \textit{intra-order hyperedge overlap}, to quantify the overlap between hyperedges of the same size and showed a wide range of values for this metric in real-world datasets. 

In this paper, we take into account both the extent of the downward closure (which marks the difference between simplicial complexes and random hypergraphs) and the overlap among hyperedges of the same order of interactions. We characterize the former by means of the \textit{inter-order hyperedge overlap} and the latter in terms of the \textit{intra-order hyperedge overlap}. To this aim, we introduce a general framework to characterize the hyperedge overlap of higher-order structures: the \textit{overlap matrix}. 
Equipped with this framework, our work aims to address three main questions: (i) the hierarchy of the importance of the two kinds of overlaps in synchronization dynamics, (ii) the distinct and combined effects of intra-order and inter-order overlaps, and (iii) the consistency of each type of overlap's influence on different dynamical systems.

The rest of the manuscript is organized as follows. In Sec.~\ref{sec:II} we introduce the overlap matrix.
In Sec.~\ref{sec:III} we illustrate the model for synchronization dynamics and highlight the significance of the spectrum of eigenvalues of the Laplacian matrix.  
Thereafter, in Sec.~\ref{sec:IV} we analyze the effect of hyperedge overlap on synchronization dynamics of coupled nonlinear oscillators. In particular, in Sec.~\ref{subsec:IVA} we consider a general scenario with $2$-body, $3$-body and $4$-body interactions and analyze the stability of the synchronous state for structures where the elements of the overlap matrix take values at the extremes of their range of definition. Our findings indicate that an increase in the overlap leads to a decrease in synchronizability. Furthermore, we elucidate a hierarchy of importance among the two types of overlap.  Finally, in Secs.~\ref{subsec:IVB} and~\ref{subsec:IVC}, we consider the simplest case of a structure with non trivial inter-order and intra-order hyperedge overlap, namely a structure with interactions up to $3-$body. We observe that both overlaps significantly influence the spectrum of eigenvalues of the Laplacian matrix, and have a qualitatively distinct impact on the stability of synchronization. In particular, we find that, for a given order, there is a close relation between intra-order hyperedge overlap and the connectedness of the set of hyperedges. Finally in Sec.~\ref{sec:V} we round off the article by discussing the main findings and some possible directions for future research.


\section{Hyperedge overlap}
\label{sec:II}

In a complex system, interactions may involve two or more units. Two-body or pairwise interactions can be modelled by networks, whereas, to represent multi-body or higher-order interactions, hypergraphs and simplicial complexes can be used. A hypergraph $H=(\mathcal{N},\mathcal{E})$ is defined as a pair of two sets: the set $\mathcal{N}$ that is composed of $N = | \mathcal{N}|$ nodes, and the set $\mathcal{E}$ that contains a number $E = | \mathcal{E}|$ of hyperedges. A hyperedge $e \in \mathcal{E}$ of order $m$, shortly an $m$-hyperedge, is defined as a subset of $m+1$ nodes in $\mathcal{N}$ with $m=|e|-1$. Thus, for each order $m$, the $m$-hyperedges capture the interactions of groups of size $m+1$. This way, pairwise interactions involving two nodes correspond to hyperedges of order 1, interactions between three nodes are represented by hyperedges of order 2, and so on for larger groups. A simplicial complex is a special case of hypergraph subject to the inclusion property (also known as downward closure), which means that all the possible subgroups of nodes appearing in a hyperedge are also connected by hyperedges of the structure.

\begin{figure*}[t!]
\centering\includegraphics[width=0.92\linewidth]{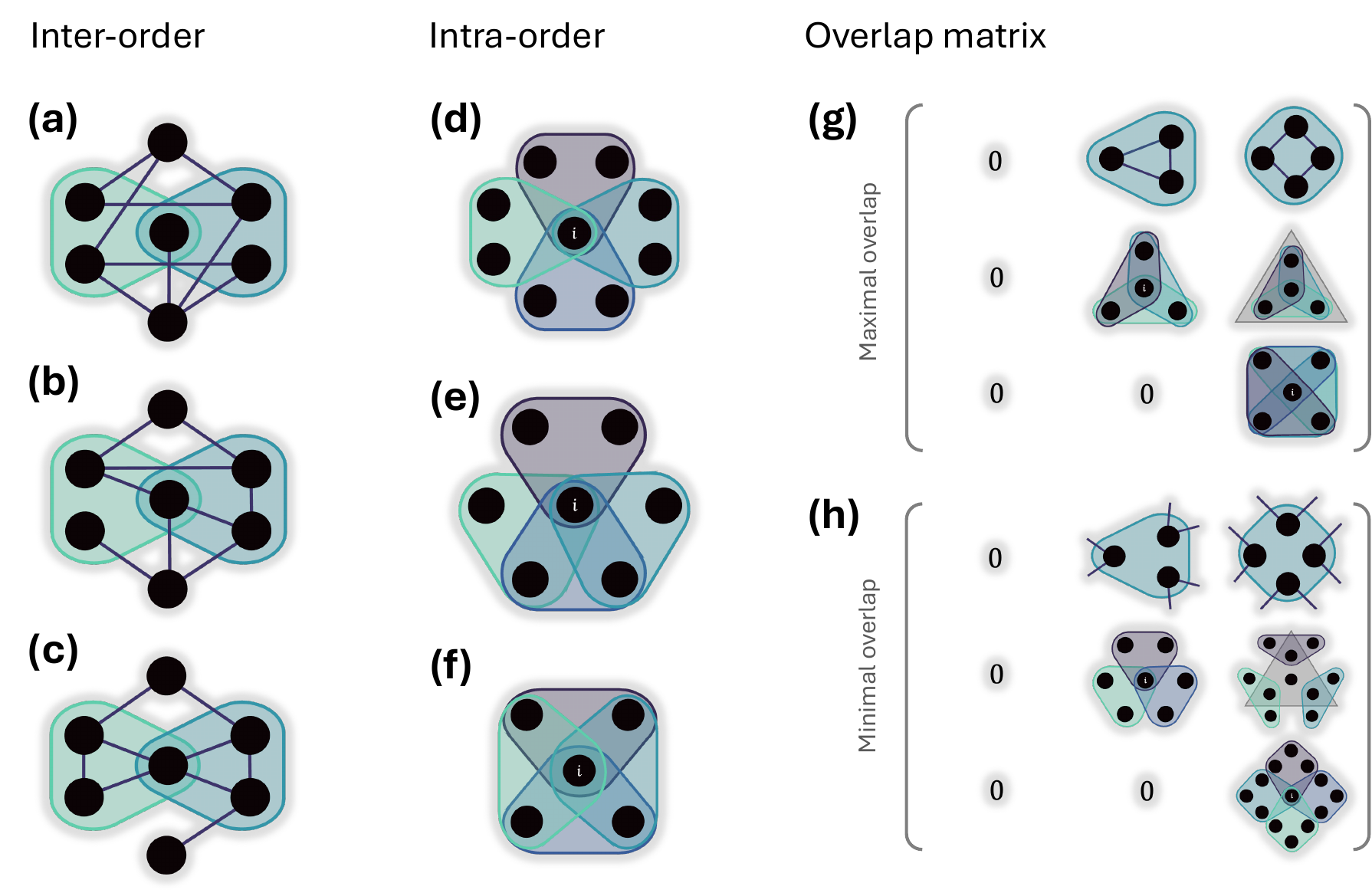}
\vspace{1cm}\caption{\textbf{Inter-order and intra-order hyperedge overlap as fundamental components of the overlap matrix.} (a)-(c): Three configurations of seven nodes, nine $1-$hyperedges and two $2-$hyperedges displaying different levels of inter-order hyperedge overlap: (a) $\mathcal{I}^{(1,2)}=0$; (b) $\mathcal{I}^{(1,2)}=0.5$; (c) $\mathcal{I}^{(1,2)}=1$. (d)-(f): Three configurations with four $2-$hyperedges displaying different levels of the intra-order hyperedge overlap for node $i$ ($k^{(2)}_i=4$): (d) $T_i^{(2)}=0$; (e) $T_i^{(2)}=0.5$; (f) $T_i^{(2)}=1$. 
(g)-(h): Schematic representation of the overlap matrix for hypergraphs with $M=3$. Panel (g) shows a configuration with maximal overlap and panel (h) shows a configuration with minimal overlap.
}
\label{fig:P1}
\end{figure*}

\subsection{Characterizing nodes and their interactions}

To characterize the microscopical organization of hypergraphs, we first need to introduce a few definitions. We begin by characterizing the structure of group interactions of each node $i$ by defining $\mathcal{E}_i^{(m)}$ as the set of hyperedges $e$ of order $m$ to which node $i$ belongs. The number of hyperedges of order $m$ to which $i$ belongs to is the so-called generalized degree of node $i$, namely $k_i^{(m)}$, and corresponds to the cardinality of $\mathcal{E}_i^{(m)}$, that is, $k_i^{(m)} = |\mathcal{E}_i^{(m)}|$~\cite{courtney2016generalized}. 

Next, in order to encode the information about the interactions at all orders, we consider a set of adjacency tensors $\mathrm{A}^{(m)}$ with $m=1,\ldots,M$, where $M$ is the highest order of group interactions in the structure. This way, we have that $a^{(m)}_{i,j_1\ldots, j_m}=1$ if there exists an $m$-hyperedge that contains nodes $i,j_1\ldots, j_m$, and $a^{(m)}_{i,j_1\ldots, j_m}=0$ otherwise. For each order $m$ we also consider a generalized Laplacian matrix $L^{(m)}$~\cite{lucas2020multiorder} with entries given by:
\begin{equation}
    l_{ij}^{(m)}=mk_i^{(m)}\delta_{ij}-b_{ij}^{(m)}.
    \label{eq:laplacian}
\end{equation}
\noindent where $b_{ij}^{(m)}$ represents the number of connections of order $m$ in which both $i$ and $j$ appear, that is:
\begin{equation}
b_{ij}^{(m)}=\frac{1}{(m-1)!}\sum\limits_{j_2,...,j_m}^Na^{(m)}_{i,j_1,j_2,...,j_m}\;.
    \label{eq:laplacian2}
\end{equation}


\subsection{Characterizing the structure of hyperedges}

In classical graphs, nodes can be connected only in pairs, via links (1-hyperedges). Thus, if, as usual, there are no multiple links connecting the same pair of nodes, each link of a node connects it with a different neighbor. However, when a structure has higher-order interactions, a node $i$ may be connected to the same node $j$ via different hyperedges. Quantifying the number of repeated neighbors among hyperedges, namely their {\em degree of overlap}, is fundamental to understand the microscopic organization of different structures. 

Random hypergraphs and simplicial complexes~\cite{battiston2020networks} exemplify the variability in the degree of overlap between hyperedges of different orders. In fact, in random hypergraphs, hyperedges are uncorrelated, meaning that in the thermodinamic limit, $N\rightarrow\infty$, the probability that two nodes share more than one hyperedge becomes negligible. Contrarily, in simplicial complexes, where the inclusion property holds, when a set of nodes conforms a hyperedge, all possible subsets of nodes are also linked by hyperedges, and so the overlap is large.

Repeated neighbors can be found either considering two hyperedges of the same order or of different orders. The two cases are dealt with different measures, in the following indicated as inter-order and intra-order hyperedge overlap. 

\subsubsection{Inter-order hyperedge overlap}

To quantify the extent of overlap between interactions of order $m$ and those of order $n$, with $m<n$, we consider the set of possible $m-$cliques within the $n$-hyperedges, denoted as $\mathcal{F}(\mathbfcal{E}^{(n)})$. Then, we count how many of those $m$-cliques correspond to actual $m-$hyperedges of the structure, i.e. $|\mathbfcal{E}^{(m)}\cap \mathcal{F}(\mathbfcal{E}^{(n)})|$. 
At this point, we define the inter-order hyperedge overlap $\mathcal{I}^{(m,n)}$ as the fraction of existing $m$-cliques over the number of possible $m-$cliques, namely:
\begin{equation}
\mathcal{I}^{(m,n)}=\frac{|\mathbfcal{E}^{(m)}\cap \mathcal{F}(\mathbfcal{E}^{(n)})|}{|\mathcal{F}(\mathbfcal{E}^{(n)})|},
    \label{eq:inter-order-overlap}
\end{equation}
The former expression spans from $\mathcal{I}^{(m,n)}=0$ (when there are no $m-$hyperedges that correspond to one of the possible $m-$cliques in the set of $n$-hyperedges) to $\mathcal{I}^{(m,n)}=1$ (when all the possible $m-$cliques within the $n$-hyperedges are $m$-hyperedges). Note that in case $m>n$, Eq. (\ref{eq:inter-order-overlap}) becomes $\mathcal{I}^{(m,n)}=0$.

As an example, Fig.~\ref{fig:P1}(a)-(c) shows three different configurations for a hypergraph of seven nodes with nine 1-hyperedges and two 2-hyperedges. Here, the extent of the inter-order overlap between 1-hyperedges and 2-hyperedges, i.e, $m=1$ and $n=2$, varies from the fully non-overlapping case shown in panel (a) to the extreme case of maximal hyperedge overlap shown in panel (c). In particular, we have in Fig.~\ref{fig:P1}(a) $|\mathcal{F}(\mathbfcal{E}^{(2)})|=6$ and  $|\mathbfcal{E}^{(1)}\cap \mathcal{F}(\mathbfcal{E}^{(2)})|=0$, resulting in $\mathcal{I}^{(1,2)}=0$. In Fig.~\ref{fig:P1}(b) the inter-order hyperedge overlap increases to $\mathcal{I}^{(1,2)}=0.5$ since $|\mathbfcal{E}^{(1)}\cap \mathcal{F}(\mathbfcal{E}^{(2)})|=3$. Finally,
in Fig.~\ref{fig:P1}(c) the maximal overlap ($\mathcal{I}^{(1,2)}=1$) is obtained considering that $|\mathbfcal{E}^{(1)}\cap \mathcal{F}(\mathbfcal{E}^{(2)})|=|\mathcal{F}(\mathbfcal{E}^{(2)})|=6$.
%
For the sake of illustration, the example of Fig.~\ref{fig:P1} refers to a structure having only 1-hyperedges and 2-hyperedges. 

\subsubsection{Intra-order hyperedge overlap}

To take into account the case where repeated nodes appear in hyperedges of the same order, we consider the local intra-order hyperedge overlap defined in~\cite{malizia2023hyperedge}: 
\begin{equation}
    T_i^{(m)} = 1-\dfrac{ \mathcal{S}_i^{(m)} - {\mathcal{{S}}_i^{(m),-}}}{\mathcal{{S}}_i^{(m),+} -\mathcal{{S}}_i^{(m),-}},
    \label{eq:Ti}
\end{equation}
where $\mathcal{S}_i^{(m)}$ represents the number of unique neighbors of a node $i$ that are found in its $k_i^{(m)}$ hyperedges of order $m\geq2$, and $\mathcal{{S}}_i^{(m),-}$ ($\mathcal{{S}}_i^{(m),+}$) indicates the minimum (maximum) possible value for the number of such nodes (the expressions for $\mathcal{{S}}_i^{(m),-}$ and $\mathcal{{S}}_i^{(m),+}$ are analytically derived in \cite{malizia2023hyperedge}). The quantity $T_i^{(m)}$ spans between $T_i^{(m)}=0$ indicating that there is minimum overlap between the $m-$hyperedges to which node $i$ belongs, and $T_i^{(m)}=1$ corresponding to the maximum overlap. 

As an example, in Fig.~\ref{fig:P1}(d)-(f) three configurations where the node $i$ has the same number of $2-$hyperedges ($k_i^{(2)}=4$) are shown. In Fig.~\ref{fig:P1}(d) the overlap between the hyperedges is minimum, since ${S}_i^{(2)}={S}_i^{(2),+}=8$. In Fig.~\ref{fig:P1}(e) the node intra-order hyperedge overlap is, instead, $T_i^{(m)}=0.5$ as ${S}_i^{(2)}=6$, ${S}_i^{(2),+}=8$, and ${S}_i^{(2),-}=4$. Finally, the full overlap case is depicted in Fig.~\ref{fig:P1}(f), where ${S}_i^{(2)}={S}_i^{(2),-}=4$.

To assess the level of intra-order hyperedge overlap across the entire hypergraph, we calculate the weighted average of $T_i^{(m)}$ throughout all nodes:
\begin{equation}
    \mathcal{T}^{(m)}= \frac{\sum_ik_i^{(m)}T_i^{(m)}}{\sum_ik_i^{(m)}}.
    \label{eq:global_overlapness}
\end{equation}

\subsubsection{The overlap matrix}

To characterize the hyperedge overlap in an hypergraph with interactions up to order $M$, we should consider all the $M-1$ measures of intra-order overlap, and all the $M(M-1)/2$ measures of inter-order overlap. For a compact representation of such measures, we introduce the overlap matrix $\mathcal{O}=\{o^{(m,n)}\}$ that embodies all the possible types of overlap of different order, as follows:
\begin{equation}
\mathcal{O} =
\begin{pmatrix}
0 & \mathcal{I}^{(1,2)} & \mathcal{I}^{(1,3)} & \dots & \mathcal{I}^{(1,M)} \\
\mathcal{I}^{(2,1)} & \mathcal{T}^{(2)} & \mathcal{I}^{(2,3)} & \dots & \mathcal{I}^{(2,M)} \\
\mathcal{I}^{(3,1)} & \mathcal{I}^{(3,2)} & \mathcal{T}^{(3)} & \dots & \mathcal{I}^{(3,M)} \\
\vdots & \vdots & \vdots & \ddots & \vdots \\
\mathcal{I}^{(M,1)} & \mathcal{I}^{(M,2)} & \mathcal{I}^{(M,3)} & \dots & \mathcal{T}^{(M)} \\
\end{pmatrix}.
\label{eq:overlap_matrix}
\end{equation}
Note that since, as mentioned above, $T_i^{(m)}$ is only defined for $m\geq 2$ we have set the $(1,1)$ entry of the overlap matrix equal to $0$ in order to capture that 2-body (pairwise) interactions do not overlap at all.  Moreover, the overlap matrix is an upper triangular matrix, since $\mathcal{I}^{(m,n)}=0$ in case $m>n$.

\medskip

\begin{figure*}[t!]
\centering\includegraphics[width=0.85\linewidth]{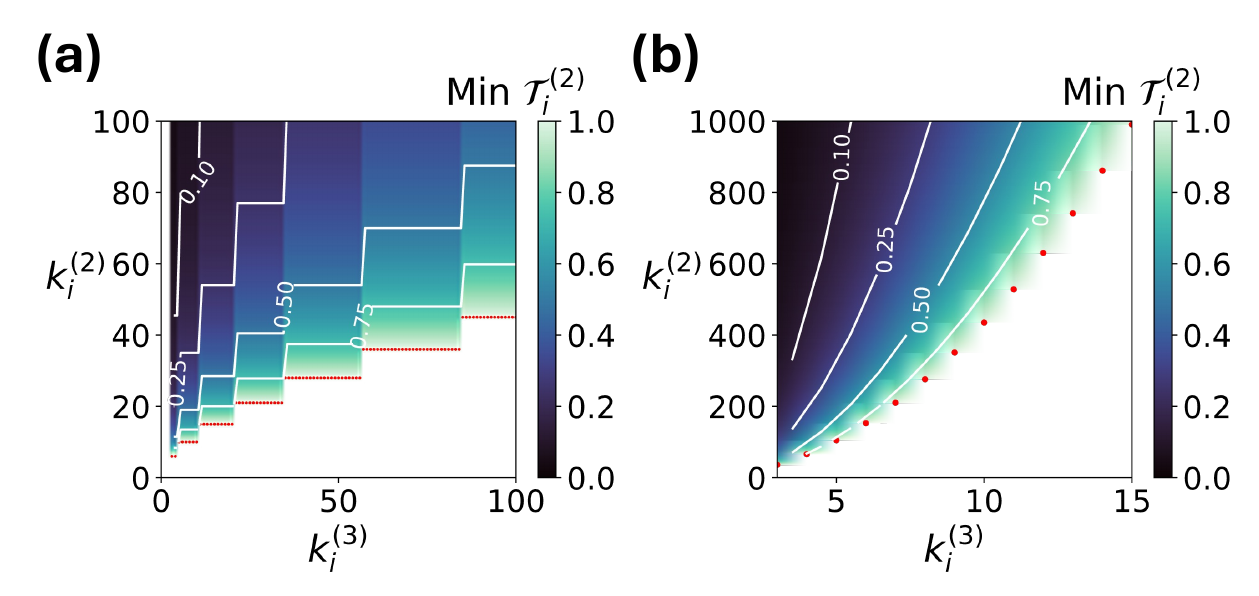}
\vspace{1cm}\caption{\textbf{Constraints between elements of the hyperedge overlap matrix.} Minimum $T^{(2)}_i$ of a node $i$ with generalized connectivity $k^{(2)}_i$ and $k^{(3)}_i$ provided $\mathcal{I}^{(2,3)}=0$. Panel (a) shows the case where $\mathcal{T}^{(3)}=1$ and panel (b) shows the case where $\mathcal{T}^{(3)}=0$. The red dots represent the minimum $k^{(2)}_i$ required to have $\mathcal{I}^{(2,3)}=0$.}
\label{fig:P2}
\end{figure*}

In Fig.~\ref{fig:P1}(g)-(h), for the case $M=3$, we exemplify the configurations yielding maximal (panel (g)) and minimal (panel (h)) values of the entries of the overlap matrix. We note that, as the overlap measures are not independent each other, not all extreme values can be reached. Specifically, when there is a non-zero inter-order overlap with $m$- and $n$-hyperedges, then for the set of $m$-hyperedges we can never find $\mathcal{T}^{(m)}=0$. In particular, in the case when $\mathcal{I}^{(m,n)}=1$, the number of unique neighbors that a node $i$ can have through hyperedges of order $m$ ($S_i^{(m)}$) depends on the number of unique neighbors that a node $i$ has through hyperedges of order $n$ ($S_i^{(n)}$) as: 
\begin{equation}
S_i^{(m)}=S_i^{(n)}+m\left(k_i^{(m)}-X\right)\;,
\end{equation} 
with $X$ being the number of different interactions of order $m$ that the node $i$ can have with the neighbors through hyperedges of order $n$, $X=$\(\binom{\;\;S_i^{(n)}}{m}\).  Therefore, the minimum intra-order hyperedge overlap of a node $i$ with $S_i^{(n)}$ unique neighbors through hyperedges of order $n$ (called $T_i^{(m)}|_{k_i^{(n)}}$) is given by:
\begin{equation}
    T_i^{(m)}|_{k_i^{(n)}} = 1 - \dfrac{ S_i^{(n)} + m \left(k_i^{(m)} - \binom{\;\;S_i^{(n)}}{m} \right) - \mathcal{S}_i^{(m),-}}{\mathcal{S}_i^{(m),+} - \mathcal{S}_i^{(m),-}}.
    \label{eq:Ti_min}
\end{equation}

Note also that there is another interdependence between the overlap measures constraining the values that it is possible to obtain. In fact, to reach $\mathcal{I}^{(m,n)}=1$, one must have that $k_i^{(m)}\geq X$. To illustrate this, let us consider $m=2$ and $m=3$, and derive the analytical expressions for $T^{(3)}=0$ and $T^{(3)}=1$. In the case of minimum intra-order hyperedge overlap, we use in Eq.~(\ref{eq:Ti_min}) the fact that $S_i^{(3)}=S_i^{(3,-)}=\lceil y^{(3)}\rceil$ \cite{malizia2023hyperedge}, where $y^{(3)}$ is the solution of $y^{(3)}\left( y^{(3)}-1\right)\left( y^{(3)}-2\right)=6k_i^{(3)}$. In the case of maximum intra-order hyperedge overlap, instead, in Eq.~(\ref{eq:Ti_min}) we consider that $S_i^{(3)}=S_i^{(3,-)}=3k_i^{(3)}$. The outcome of this analysis is illustrated in Fig.~\ref{fig:P2}(a)-(b). The red dots indicate the minimum $2$-hyperedge connectivity required to have $\mathcal{I}^{(m,n)}=1$ for a given $k_i^{(3)}$, and the color code represents the minimum possible value of $2$-hyperedge intra-order overlap in terms of both generalized connectivities ($k_i^{(2)}$, $k_i^{(3)}$). Let us notice that the minimum intra-order hyperedge overlap of $3$-hyperedges needs larger values of $2$-hyperedges connectivity than in the case of maximum intra-order overlap. These constraints become increasingly complex with the introduction of additional orders of interaction.

\section{Model of dynamical systems coupled through a hypergraph}
\label{sec:III}

Now we turn our attention to the model of dynamical systems coupled through hypergraphs, focusing on the stability analysis of synchronous (homogeneous) solutions. Let us start by considering the following generic system of $N$ oscillators coupled via $m$-body interactions with $m = 1,\ldots,M $~\cite{gambuzza2021stability}:
\begin{align}
& \dot{\mathbf{x}}_i = \mathbf{f}(\mathbf{x}_i)\nonumber \\ 
& \qquad +\sum_{m=1}^M\sigma^{(m)}
\sum\limits_{j_1,..,j_m} a_{ij_1...j_m}^{(m)}\textbf{g}^{(m)}(\textbf{x}_i,\textbf{x}_{j_1},\ldots,\textbf{x}_{j_m})\;,
\label{eq:generalcase}
\end{align}
where $\textbf{x}_i$ is the $n-$dimensional state vector associated to each node $i$, and $\mathbf{f}(\textbf{x}_i)$ describes the local dynamics, assumed identical for all units. The functions $\textbf{g}^{(m)}$ capture the coupling mechanisms between units at each order $m$ of interaction, whereas the constant parameters $\sigma^{(m)},\; m=1,\ldots,M$ represent the coupling strengths associated to these interactions, and $a_{ij_1...j_m}^{(m)}$ are the entries of the corresponding adjacency tensor $A^{(m)}$.

To guarantee the existence and invariance of a synchronous solution of the type $\textbf{x}_1=...=\textbf{x}_N=\textbf{x}^s$, we assume that the coupling functions are non-invasive, i.e.:
\begin{equation}
\textbf{g}^{(m)}(\textbf{x},\textbf{x},...,\textbf{x})\equiv 0,\; m=1,\ldots,M. 
\end{equation}

However, this is not enough for the stability of the synchronous solution that requires further conditions, as it depends on the interplay between the local dynamics and the structure of interactions between units. To unveil this interplay and determine the conditions for synchronization stability, as in~\cite{gambuzza2021stability} 
we start by considering a small perturbation around the synchronous state, i.e., for each node $i$ we consider $\delta\textbf{x}_i=\textbf{x}_i-\textbf{x}^s$. When interactions between units are exclusively pairwise, the former perturbations can be expressed as a linear combination of the eigenvectors of the network Laplacian matrix $L^{(1)}$, denoted as $\boldsymbol{\eta}_i$. The dynamics of the component $\boldsymbol{\eta}_1$ (corresponding to the zero eigenvalue $\lambda_1=0$) is responsible for the motion along the synchronous manifold, while the dynamics of the remaining components corresponding to $\boldsymbol{\eta}_i$ ($i=2,...,N$) represent the modes transverse to the synchronization manifold. The linear stability of the synchronous state requires that the dynamics of these $N-1$ transverse modes damp out. This condition can be checked by studying the maximum Lyapunov exponent associated to the transverse modes. In the presence of higher-order interactions, the transverse modes are usually intertwined, and the study of synchronization stability has to be performed on a set of linear equations in a number equal to the size of the structure (the number of nodes). Nevertheless, for a large class of coupling functions, the dynamics of the transverse modes can be decoupled, allowing the calculation of the stability conditions from a single parameter variational equation, having the same dimension of the dynamical system at work in each (isolated) node. This decoupling can be achieved by considering the class of diffusive-like coupling functions, for which
\begin{align}
& \textbf{g}^{(m)}(\textbf{x}_i,\textbf{x}_{j_1},...,\textbf{x}_{j_m})\nonumber \\ 
& \qquad\qquad = \textbf{h}^{(m)}(\textbf{x}_{j_1},...,\textbf{x}_{j_m}) - \textbf{h}^{(m)}(\textbf{x}_i,...,\textbf{x}_i)\;,
\label{eq:gmhm}
\end{align}
under the assumption of natural couplings~\cite{gambuzza2021stability}, i.e. 
\begin{equation}
\textbf{h}^{(m)}(\textbf{x},...,\textbf{x})=\textbf{h}^{(1)}(\textbf{x})\; , m=1,\ldots,M.
\end{equation}
Under these conditions, the parametric equation characterizing the dynamics of the transverse modes, and therefore the synchronization stability of system~\eqref{eq:generalcase}, reads:
\begin{equation}
    \dot{\boldsymbol{\eta}}=\left[J\textbf{f}  (\textbf{x}^s)-\alpha J \textbf{h}^{(1)}(\textbf{x}^s)\right]\boldsymbol{\eta}.
    \label{eq:MSF}
\end{equation}
where $J\textbf{f}(\textbf{x}^s)$ and $J\textbf{h}^{(1)}(\textbf{x}^s)$ represents the Jacobian matrix of the local dynamics and the coupling functions, respectively, both calculated around the synchronous solution $\textbf{x}^s$. From this equation we compute the maximum Lyapunov exponent as a function of the parameter $\alpha$, namely the master stability function $\Lambda_{\text{max}}=\Lambda_{\text{max}}(\alpha)$ characterizing the stability of synchronization \cite{pecora1998master}. In more detail, synchronization stability requires that the master stability function takes negative values in correspondence of the points $\alpha=\{\lambda_2(\bar L),\ldots,\lambda_N(\bar L)\}$, where $\bar L$ is the effective Laplacian~\cite{gambuzza2021stability,lucas2020multiorder}, defined as:
\begin{equation}
    \bar L = \sum_{m=1}^M\sigma^{(m)} L^{(m)}.
    \label{eq:effective_laplacian}
\end{equation}
The eigenvalues $\lambda_n(\bar L)$, with $n=1,\ldots,N$, of the effective Laplacian matrix (which is, by construction, positive semi-definite) are labeled throughout the manuscript in ascending order of magnitude, i.e. $0=\lambda_1(\bar L)<\lambda_2(\bar L)\leq \ldots \leq \lambda_N(\bar L)$.


\begin{figure*}[t!]
\centering\includegraphics[width=1\linewidth]{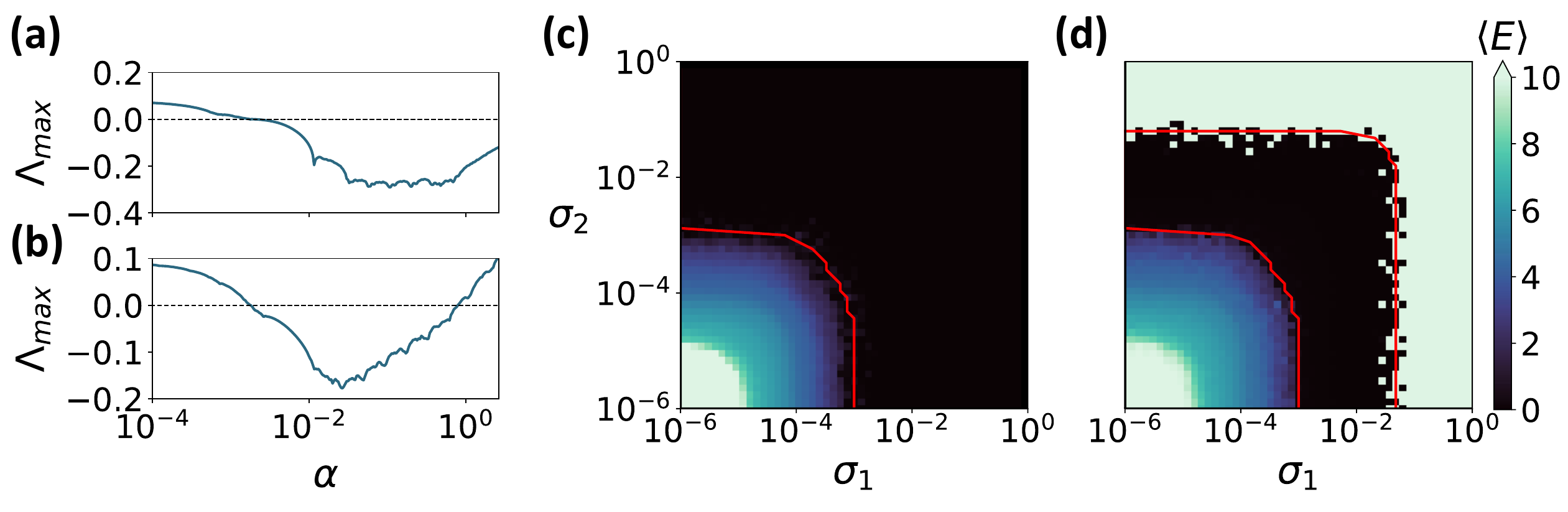}
\vspace{1cm}\caption{\textbf{Synchronization in a hypernetwork of Rössler oscillators} (a)-(b): Master stability function calculated for different coupling functions. (a) Class II system given $\textbf{h}^{(1)}(\textbf{x}_j)=[0,y_j^3,0]^T$ and $\textbf{h}^{(2)}(\textbf{x}_j,\textbf{x}_k)=[0,y_j^2y_k,0]^T$. (b) Class III system given $\textbf{h}^{(1)}(\textbf{x}_j)=[x_j^3,0,0]^T$ and $\textbf{h}^{(2)}(\textbf{x}_j,\textbf{x}_k)=[x_j^2x_k,0,0]^T$. (c)-(d): Phase diagram of a system of coupled Rössler oscillators, displaying the synchronization error in terms of the $1-$hyperedge and $2-$hyperedge coupling strength $E(\sigma^{(1)},\sigma^{(2)})$. (c) Class II system given in accordance with (a). (d) Class III system in accordance with (b). The red lines correspond to the theoretical predictions of the synchronization threshold given by the MSF approach. In both panels the structure has $N=100$, $k^{(1)}=6$, $k^{(2)}=3$, and has zero hyperedge overlap ($\mathcal{I}^{(1,2)}=0$, $T^{(2)}=0$). }
\label{fig:P3}
\end{figure*}


The properties of the master stability function $\Lambda_{\text{max}} = \Lambda_{\text{max}}(\alpha)$ determine the class of the system of coupled oscillators. Specifically, as discussed in~\cite{boccaletti2006complex,arenas2008synchronization,gallo2022synchronization} one can identify three classes of systems:

\begin{itemize}
    \item \textit{Class I systems}: $\Lambda_{\text{max}}$ is always positive, regardless of $\alpha$. Consequently, the system remains incoherent.
    \item \textit{Class II systems}: As shown in Fig.~\ref{fig:P3}(a), $\Lambda_{\text{max}}$ crosses the $\alpha$-axis once at $\alpha_c$. For synchronization stability, it is required that $\lambda_2(\bar{L}) > \alpha_c$.
    \item \textit{Class III systems}: As shown in Fig.~\ref{fig:P3}(b), the master stability function crosses the $\alpha$-axis at two points, $\alpha_1$ and $\alpha_2$. In this case, the stability region is bounded by two conditions: $\lambda_2(\bar{L}) > \alpha_1$ and $\lambda_N(\bar{L}) < \alpha_2$.
\end{itemize}

In this paper, we consider a system of $N$ coupled R\"ossler oscillators that, depending on the coupling functions used in \eqref{eq:generalcase}, can yield either a class II or class III system.
The individual dynamics of a node $i$, $\textbf{f}(\textbf{x}_i)$, with $\textbf{x}_i=(x_i,y_i,z_i)$ are described by the following systems of equations:
\begin{equation}
\begin{array}{lll}
\dot{x}_i & = & -y_i - z_i\;,\\
\dot{y}_i & = & x_i + a y_i\;,\\
\dot{z}_i & = & b + z_i (x_i - c)\;,
\label{eq:Iz}
\end{array}
\end{equation}
where the parameters are fixed to $a = 0.2$, $b = 0.2$, and $c = 9$, so that the isolated dynamics is chaotic. A class II system is obtained for instance when  $\textbf{g}^{(1)}(\textbf{x}_i,\textbf{x}_j)=\textbf{h}^{(1)}(\textbf{x}_j)=[0,y_j^3,0]^T$ and $\textbf{g}^{(2)}(\textbf{x}_i,\textbf{x}_j,\textbf{x}_k)=\textbf{h}^{(2)}(\textbf{x}_j,\textbf{x}_k)=[0,y_j^2y_k,0]^T$, that is, when the nodes are coupled by their second component $y$:
\begin{equation}
\label{eq:classIIsystem}
\begin{array}{lll}
\dot{x}_{i} & = & -y_{i} -z_{i}\;,
\\
\dot{y}_{i} & = & x_{i} + a y_{i} +\sigma_1\sum\limits_{j=1}^N a_{ij}^{(1)}(y_j^3-y_i^3)\\
& & +\sigma_2\sum\limits_{j=1}^N\sum\limits_{k=1}^N a_{ijk}^{(2)}(y_j^2y_k-y_i^3)\;,
\\
\dot{z}_{i} & = & b + z_{i} (x_{i} -c)\;,
\end{array}
\end{equation}
Conversely, a class III system is obtained for instance when $\textbf{g}^{(1)}(\textbf{x}_i,\textbf{x}_j)=\textbf{h}^{(1)}(\textbf{x}_j)=[x_j^3,0,0]^T$ and $\textbf{g}^{(2)}(\textbf{x}_i,\textbf{x}_j,\textbf{x}_k)=\textbf{h}^{(2)}(\textbf{x}_j,\textbf{x}_k)=[x_j^2x_k,0,0]^T$, such that the oscillators are coupled through the variable $x$:
\begin{equation}
\label{eq:caseIIIsystem}
\begin{array}{lll}
\dot{x}_{i} & = & -y_{i} -z_{i} +\sigma_1\sum\limits_{j=1}^N a_{ij}^{(1)}(x_j^3-x_i^3)\\
& & +\sigma_2\sum\limits_{j=1}^N\sum\limits_{k=1}^N a_{ijk}^{(2)}(x_j^2x_k-x_i^3)\;,
\\
\dot{y}_{i} & = & x_{i} + a y_{i}\;,
\\
\dot{z}_{i} & = & b + z_{i} (x_{i} -c)\;,
\end{array}
\end{equation}

To monitor synchronization among the chaotic units we use the following synchronization error:
\begin{equation}
    E(t)=\left(\frac{1}{N(N-1)}\sum_{i,j=1}^N||\textbf{x}_j-\textbf{x}_i||^2\right)^{\frac{1}{2}},
    \label{fig:synch-error}
\end{equation}
which vanishes in the case of complete synchronization, and takes large values for incoherent behavior. In particular, we let system~\eqref{eq:classIIsystem} or~\eqref{eq:caseIIIsystem}  evolve for a transient of duration $t_r=750$ and, afterwards, calculate the average of the error defined as:
\begin{equation}
\langle E\rangle=\frac{1}{T}\int_{t_r}^{t_r+T}E(t)dt\end{equation} 
for a time window of duration $T=750$.

\begin{figure}[t!]
\centering\includegraphics[width=1\linewidth]{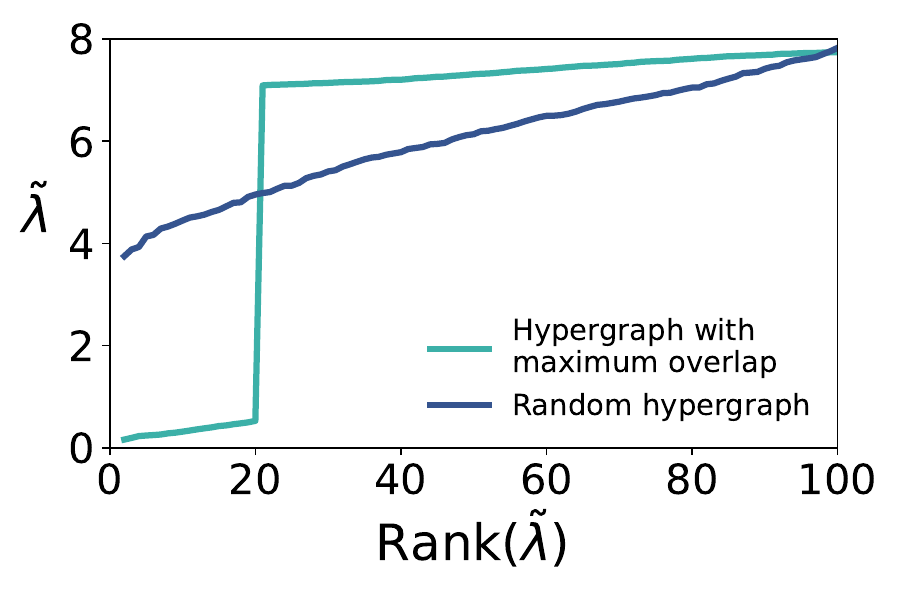}
\vspace{1cm}\caption{\textbf{Influence of the hyperedge overlap on the spectrum of the effective Laplacian.} Spectrum of the effective Laplacian for two regular higher-order structures with $N=100$, $k^{(1)}=6$, $k^{(2)}=6$ and $k^{(3)}=4$. In the maximum overlap structure $\{\mathcal{T}^{(2)},\mathcal{T}^{(3)},\mathcal{I}^{(1,2)},\mathcal{I}^{(1,3)}\mathcal{I}^{(2,3)}\}=\{1,1,1,1,1\}$, and in the random structure $\{\mathcal{T}^{(2)},\mathcal{T}^{(3)},\mathcal{I}^{(1,2)},\mathcal{I}^{(1,3)}\mathcal{I}^{(2,3)}\}\approx\{0,0,0,0,0\}$.}
\label{fig:P4}
\end{figure}

In Fig.~\ref{fig:P3}(c) and~\ref{fig:P3}(d) we illustrate an example of synchronization in a system of $N=100$ R\"ossler nodes interacting in a random hypergraph with $k^{(1)}=6$, $k^{(2)}=3$, $\mathcal{I}^{(1,2)}=0$, and $T^{(2)}=0$. In particular, the average error $\langle E\rangle$ is studied as a function of the two coupling strengths $\sigma_1$ and $\sigma_2$, that is, $\langle E\rangle\left(\sigma_1,\sigma_2\right)$. Panel (c) shows the results for the class II system, while panel (d) for the class III system. The values of the average error $\langle E\rangle\left(\sigma_1,\sigma_2\right)$ obtained from the numerical simulations are color-coded, while the red lines represent the theoretical predictions of the boundaries of the region where the synchronous solution is stable. Numerical simulations are in agreement with the theoretical predictions based on the master stability function approach.

The class II case, depicted in Fig.~\ref{fig:P3}(c), displays a region of incoherence in correspondence of low values of the two coupling strengths. As at least one of the two coupling strength is increased, we find a region where synchronization is stable. The transition from incoherent behavior to synchronization also appears in Fig.~\ref{fig:P3}(d) for the class III system (but shifted according to $\alpha_1\mapsto\alpha_c$). However, in this case, there is also another transition, observed for large values of the coupling strengths, where synchronization stability is lost and the system behaves again in an incoherent way. As expected, we note that the region of stability is unbounded for the class II system and bounded for the class III system. 

For the sake of brevity, hereafter we only analyze the class III scenario, since the class II behavior straightforwardly arises when relaxing the constraint on the last eigenvalue. In fact, every consequence of the constraint $\lambda_2(\bar{L}) > \alpha_1$ is also applicable to the constraint $\lambda_2(\bar{L}) > \alpha_c$, via a shift $\alpha_1\mapsto\alpha_c$.

Note that the significance of the second eigenvalue of the Laplacian extends beyond the dynamical systems of coupled chaotic oscillators. In the higher-order model of Kuramoto oscillators sharing identical natural frequencies, it governs the rate at which the system recovers its collective state after a perturbation \cite{lucas2020multiorder}, 
 with larger values of \(\lambda_2(\bar{L})\) resulting in faster convergence to the equilibrium state. 

In the rest of the paper, we will refer to the second eigenvalue of the effective Laplacian matrix in Eq.~\eqref{eq:effective_laplacian} as the \textit{algebraic connectivity}, generalizing the notion for systems with only pairwise interactions. Overall, the spectrum of the effective Laplacian matrix is important for characterizing the dynamics of numerous systems \cite{arenas2008synchronization}.

\begin{figure*}[t!]
\centering\includegraphics[width=1\linewidth]{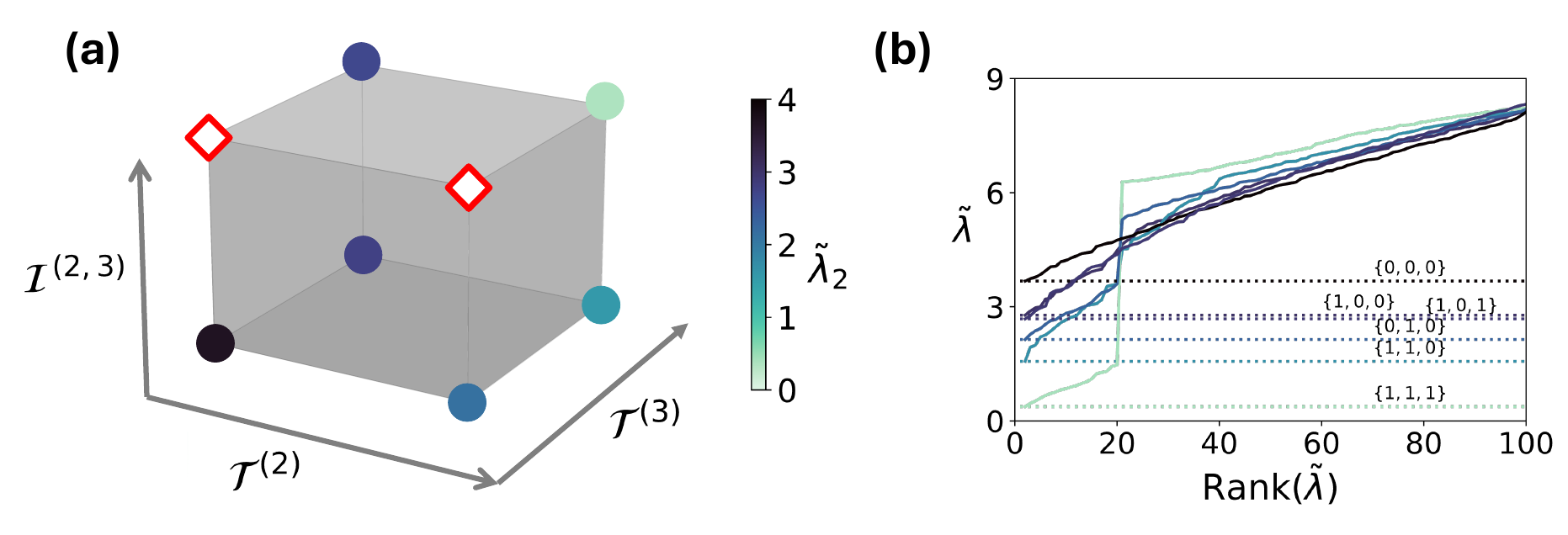}
\vspace{1cm}\caption{\textbf{Hyperedge overlap and spectrum of the effective Laplacian.}
(a) Space of parameters $\{\mathcal{T}^{(2)},\mathcal{T}^{(3)},\mathcal{I}^{(2,3)}\}$ for hypergraphs with $M=3$. In the corners, the color code represents the value of the second eigenvalue of the effective Laplacian given $\gamma=1$, namely $\tilde\lambda_2$. 
(b) Spectrum of the effective Laplacian for regular higher-order structures at the extreme positions in the $\{\mathcal{T}^{(2)},\mathcal{T}^{(3)},\mathcal{I}^{(2,3)}\}$-space.
The considered hypergraphs have $N=100$, $k^{(1)}=2$, $k^{(2)}=6$ and $k^{(3)}=4$ except for the $\{1,0,1\}$ one, which has $k^{(2)}=12$. In all cases, the hypergraphs have $\{\mathcal{I}^{(1,2)},\mathcal{I}^{(1,3)}\}\approx\{0,0\}$.}
\label{fig:P5}
\end{figure*}

\section{Results}
\label{sec:IV}
Once introduced the essential tools to carry out our analysis, in this section we focus on how the microscopic arrangement of higher-order interactions, driven by the overlap matrix defined in Eq.~\eqref{eq:overlap_matrix}, influences the stability of synchronous states.

\subsection{Effect of hyperedge overlap on the spectrum of the effective Laplacian matrix}
\label{subsec:IVA}

In this section, we investigate the impact of hyperedge overlap on synchronization stability by examining its influence on the spectrum of the effective Laplacian matrix. To this end, we consider structures with interactions up to order $M=3$, i.e. encompassing two-body, three-body and four-body interactions. Furthermore, in Eq.~(\ref{eq:effective_laplacian}) we assign equal relevance to the interactions across all orders by setting $\sigma^{(m)}=\gamma^{(m)}/k^{(m)}$ and $\gamma^{(m)}=1$ ($m=1,2,3$). To simplify our notation, we define $\tilde\lambda_n \equiv \lambda_n(\bar{L})$ for this scenario, and keep $\lambda_n(\bar L)$ for the general case of $\gamma^{(m)}\neq1$.

We begin analyzing two configurations where the overlaps take values at the extremes of the interval of definition. In Fig.~\ref{fig:P4} we compare the spectrum of the effective Laplacian matrix of a random hypergraph, displaying almost zero overplap, $o^{(m,n)}\approx 0,\;\forall\;m,n\leq3$, with that of a hypergraph designed to have maximum overlap ($o^{(m,n)}=1,\;\forall\;m,n\leq3$ but $m=n=1$). Both structures share the same number of nodes ($N=100$) and the same generalized degree distributions, specifically $k^{(1)}=6$, $k^{(2)}=6$, and $k^{(3)}=4$. Beyond the first eigenvalue that, being zero by definition, is the same in both cases, the plots reveal significant differences in the two spectrums. In more detail, we find a large discrepancy in the algebraic connectivity, $\tilde\lambda_2$, which is much larger for the random hypergraph. This corroborates the findings reported in \cite{zhang2023higher}, where it is shown that random hypergraphs facilitate synchronization to a greater extent than simplicial complexes, which, in fact, correspond to structures with $o^{(m,n)}=1,\;\forall m,n$ but $m=n=1$ (note that in \cite{zhang2023higher} a system of coupled phase-only oscillators where synchronization only depends on the algebraic connectivity is considered). Regarding the rest of the spectrum we observe that, in the random structure, the value of eigenvalues increases smoothly as the rank progresses. In contrast, in the maximum overlap scenario, the spectrum, after a first increase, undergoes an abrupt shift between the eigenvalues $\tilde\lambda_{20}$ and $\tilde\lambda_{21}$. Moreover, for $\tilde\lambda>\tilde\lambda_{20}$, the eigenvalues are larger than the corresponding ones in the random hypergaphs. This is a consequence of the mesoscale organization that is necessary to achieve a maximum intra-order overlap configuration. More specifically, to have a regular hypergraph with $k^{(2)}=6$ and $k^{(3)}=4$ and maximum intra-order hyperedge overlap for both orders $m=2$ and $m=3$, i.e., $\mathcal{T}^{(2)}=\mathcal{T}^{(3)}=1$, the nodes must be arranged on $20$ subsets of $5$ nodes, with each of these subsect being fully connected by 3- and 4-body interactions. Consequently, the $20$ smallest eigenvalues of the spectrum correspond to the first effective eigenvalue of each subset of $5$ nodes, while the next 20 ones correspond to the effective algebraic connectivity of each subset of $5$ nodes. The small value of the first (which would be zero in the absence of pairs) and the large value of the second indicate that, while synchronization stability is hindered on a global scale, there is a tendency towards local synchronization.

The two scenarios depicted in Fig.~\ref{fig:P4} illustrate configurations in which all elements in the overlap matrix~\eqref{eq:overlap_matrix} are either entirely ones or entirely zeros (except for the element (1,1)). To better illustrate the interplay between different elements of the overlap matrix, we now consider some configurations with various combinations of ones and zeros within the structure of the overlap matrix. Note that, assuming that the elements of $\mathcal{O}$ can only take binary values, there are $2^5=32$ possible different overlap matrices. Here, we focus on eight of them obtained by varying only three elements of the overlap matrix, namely, $\{\mathcal{T}^{(2)}, \mathcal{T}^{(3)}, \mathcal{I}^{(2,3)}\}$, while we keep $\mathcal{I}^{(1,2)}\approx\mathcal{I}^{(1,3)}\approx0$. These eight combinations are represented in Fig.~\ref{fig:P5}(a), at the corners of a cube. The color code is used to represent the magnitude of the algebraic connectivity $\tilde\lambda_2$. Notably, the configurations at the corners $\{0,0,1\}$ and $\{1,0,1\}$ are impossible to reach due to the aforementioned constraints (Eq.~(\ref{eq:Ti_min})) between intra-order hyperedge overlaps in case that the inter-order hyperedge overlap is maximum.

\begin{figure*}[t!]
\centering\includegraphics[width=1\linewidth]{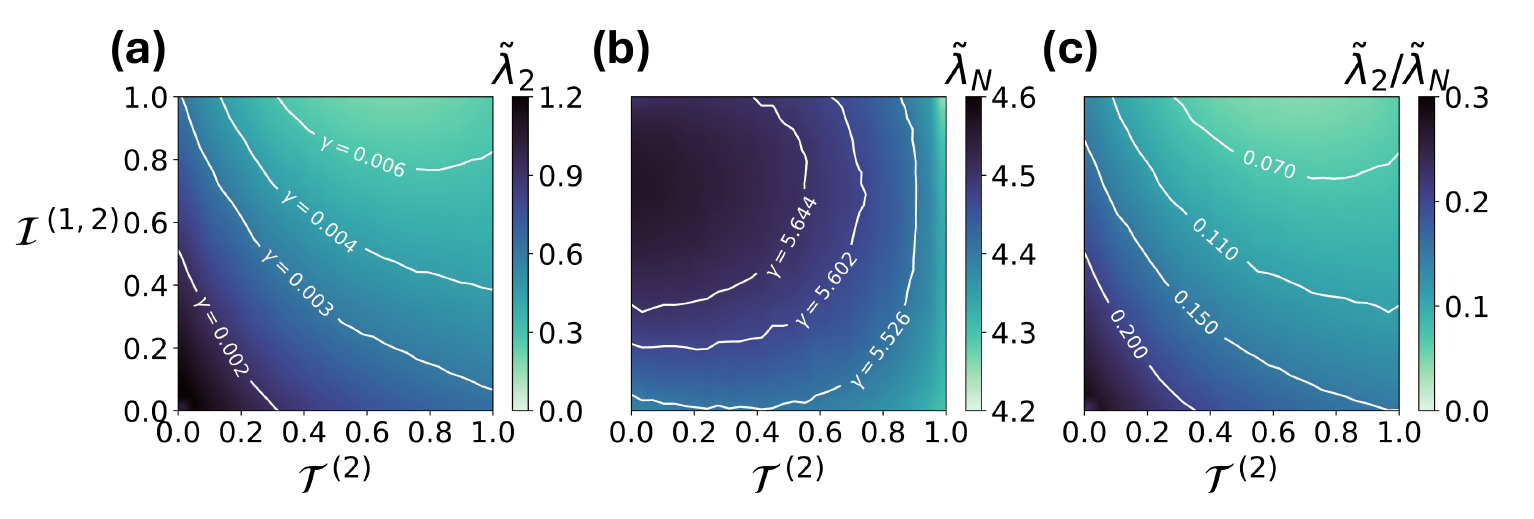}
\vspace{1cm}\caption{\textbf{Inter-order and intra-order hyperedge overlap impact the conditions for synchronization stability.} (a)-(c): Average $\tilde\lambda_2$, $\tilde\lambda_N$ and $\tilde\lambda_2/\tilde\lambda_N$ in the $\{\mathcal{T}^{(2)},\mathcal{I}^{(1,2)}\}$-space over 100 sets of structures, each with $N=100$ nodes and connectivity $k^{(1)}=6$ and $k^{(2)}=3$. The contour lines in panels (a) and (b) correspond to $\gamma\tilde\lambda_2=\alpha_1$ and $\gamma\tilde\lambda_N=\alpha_2$, respectively. 
}
\label{fig:P6}
\end{figure*}

In Fig.~\ref{fig:P5}(b), we show the spectrum of eigenvalues for the six possible configurations analyzed. As shown, the spectrum of the configurations $\{0,0,0\}$ and $\{1,1,1\}$ exhibit a behavior similar to that of the cases with minimum and maximum hyperedge overlap, respectively, shown in Fig.~\ref{fig:P4}. 
As already noted, the lowest algebraic connectivity $\tilde\lambda_2$ is found for the structure with the highest degree of overlap ($\{1,1,1\}$), followed by the configuration with $\{1,1,0\}$. In the maximum overlap structure $\{1,1,1\}$, the subsets of nodes fully-connected by $2$-hyperedges coincide with those fully-connected by $3$-hyperedges due to the downward closure. In contrast, in the structure $\{1,1,0\}$, the subsets of nodes fully-connected by $2$-hyperedges and $3$-hyperedges are intertwined, so that the stability of the synchronous state (being a collective state shared by all nodes this represents global synchronization) is fostered, but the tendency towards local synchronization is hindered. The configuration $\{0,1,0\}$ has random $3$-body interactions, unlike the former $\{1,1,0\}$. For this reason, this configuration also fosters global synchronization stability, since it leads the third lowest value of $\tilde\lambda_2$.
Furthermore, the random 3-body interactions do not interfere with 4-body interactions, which leads to a more clearly defined clustered mesoscale configuration, resulting in eigenvalues of the second group (those after the abrupt shift) larger than in the configuration $\{1,1,0\}$. Finally, the configurations  $\{1,0,0\}$ and $\{1,0,1\}$ exhibit a similar spectrum, although the former displays a larger degree of inter-order hyperedge overlap, which impairs stability of synchronization.

\subsection{Isolated and combined effect of inter- and intra-order hyperedge overlaps}
\label{subsec:IVB}

In this section, we study the simplest scenario where inter- and intra-order hyperedge overlaps can be defined, namely hypergraphs with 2- and 3-body interactions ($M=2$). In this case, the overlap matrix is $2\times 2$ and, thus, the only control parameters are the inter-order hyperedge overlap $\mathcal{T}^{(2)}$ and the intra-order hyperedge overlap $\mathcal{I}^{(1,2)}$. Here we want to conveniently adjust these two parameters to assess their isolated and combined effects on the spectrum of the effective Laplacian. This approach allows for a more comprehensive analysis compared to the limited set of configurations with distinct overlap values examined in Sec.~\ref{subsec:IVA}. In fact, in this case we are able to tune the parameters to cover the full range of possible values rather than considering only binary values of overlap. To this aim, since random hypergraphs models do not allow full control of the values of hyperedge overlap, we need a synthetic model covering the full range of values $[0,1]$ for both $\mathcal{T}^{(2)}$ and $\mathcal{I}^{(1,2)}$ (detailed information on the construction of this synthetic hypernetwork can be found in Appendix A). Note that, also in this case, to ensure equal relevance across different orders, we set $\sigma^{(m)} = \gamma^{(m)}/k^{(m)}$ and $\gamma^{(m)} = 1$ for $m=1,2$. 
Finally, we mention that our results are averaged over a number of 100 hypernetworks for each pair of values in the $\{\mathcal{T}^{(2)},\mathcal{I}^{(1,2)}\}$-space. Each hypernetwork has $N=100$ nodes, and generalized degrees $k^{(1)} = 6$ and $k^{(2)} = 3$.

\begin{figure*}[t!]
\centering\includegraphics[width=0.9\linewidth]{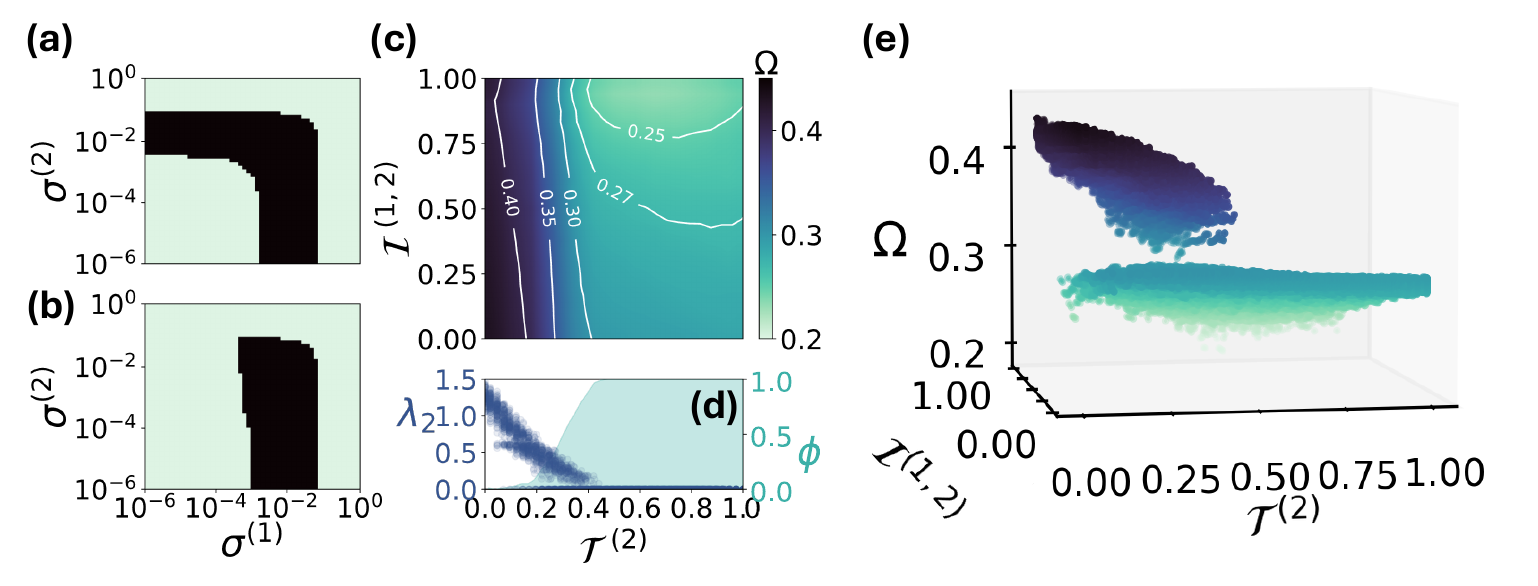}
\vspace{1cm}\caption{\textbf{Intra-order hyperedge overlap may change the shape of the region of synchronization stability.} 
(a)-(b) Region of synchronization stability (black area) for
a system of Rössler oscillators with $\textbf{h}^{(1)}(\textbf{x}_j)=[x_j^3,0,0]^T$ and $\textbf{h}^{(2)}(\textbf{x}_j,\textbf{x}_k)=[x_j^2x_k,0,0]^T$. (a) Hypergraph with $\{\mathcal{T}^{(2)},\mathcal{I}^{(1,2)}\}=\{0,1\}$. (b) Hypergraph with $\{\mathcal{T}^{(2)},\mathcal{I}^{(1,2)}\}=\{1,1\}$.
(c) Average $\Omega$ in the $\{\mathcal{T}^{(2)},\mathcal{I}^{(1,2)}\}$-space over 100 sets of structures.
(d) Scatter plot (in blue) of the second eigenvalue of each of the sets of $2$-hyperedges, that is, $\lambda_2(L^{(2)})$, as a function of $\mathcal{T}^{(2)}$. Fraction $\phi$ of structures with a given $\mathcal{T}^{(2)}$ and $\lambda_2(L^{(2)})=0$ (in green).
(e) Values of $\Omega$ in the $\{\mathcal{T}^{(2)},\mathcal{I}^{(1,2)}\}$-space.
In all panels the structures have $N=100$, $k^{(1)}=6$, $k^{(2)}=3$.}
\label{fig:P7}
\end{figure*}

We begin our analysis of the influence of the hyperedge overlap on the stability of the synchronous state in these hypergraphs by showing in Fig.~\ref{fig:P6}(a) the algebraic connectivity $\tilde\lambda_2$ as a function of $\mathcal{T}^{(2)}$ and $\mathcal{I}^{(1,2)}$. 
%
%
The color code shows values of $\tilde\lambda_2$  depending almost symmetrically on both metrics, pinpointing that inter- and intra-order hyperedge overlap have an equivalent and complementary impact. Furthermore, the insights obtained from  
Fig.~\ref{fig:P4} and Fig.~\ref{fig:P5} are corroborated, since in the case when both hyperedge-overlap metrics are at their minimum value, namely when $\mathcal{T}^{(2)}=\mathcal{I}^{(1,2)}=0$, the second eigenvalue of the effective Laplacian is at its maximum.

%

As discussed in Sec.~\ref{sec:III}, for class III systems, stability of the synchronous state depends on $\lambda_2(\bar L)$ and $\lambda_N(\bar L)$. In particular, the condition for the algebraic connectivity reads $\lambda_2(\bar L)>\alpha_1$. In the case that, $\gamma^{(m)}=\gamma,\; m=1,\ldots,M$, we have that $\lambda_2(\bar L)=\gamma\tilde\lambda_2$. For this reason, in Fig.~\ref{fig:P6}(a) we show the contour lines corresponding to $\gamma\tilde\lambda_2=\alpha_1$, namely the borders between the region of stability and instability. For each $\gamma$, values of the inter- and intra-order overlap above the corresponding contour line yield $\lambda_2(\bar L)<\alpha_1$, thus indicating that the synchronous state is not stable. As expected, the stability region increases as the parameter $\gamma$ increases. The second condition for synchronization stability, namely $\lambda_N(\bar L)<\alpha_2$, is analyzed in Fig.~\ref{fig:P6}(b), showing the values of $\tilde\lambda_N$ for $\gamma=1$ as a function of $\mathcal{T}^{(2)}$ and $\mathcal{I}^{(1,2)}$, as well as the contour lines $\gamma\tilde\lambda_N=\alpha_2$ for three different values of $\gamma$. In this case, the two overlaps exert a different effect on $\tilde\lambda_N$, with large intra-order hyperedge overlap resulting in the highest values. Moreover, the worst scenario for stability of the synchronous state occurs for low values of the intra-order hyperedge overlap and medium to high values of inter-order hyperedge overlap. Nonetheless, the relative variation of $\tilde\lambda_N$ across the parameter space is less pronounced than that observed for the second eigenvalue $\tilde\lambda_2$. This consideration is confirmed by the analysis of the eigenvalue ratio $\tilde\lambda_2/\tilde\lambda_N$ shown in Fig.~\ref{fig:P6}(c), whose dependency on the hyperedege overlap is similar to the one observed for the second eigenvalue $\tilde\lambda_2$ in Fig.~\ref{fig:P6}(a).

\begin{figure*}[t!]
\centering\includegraphics[width=0.85\linewidth]{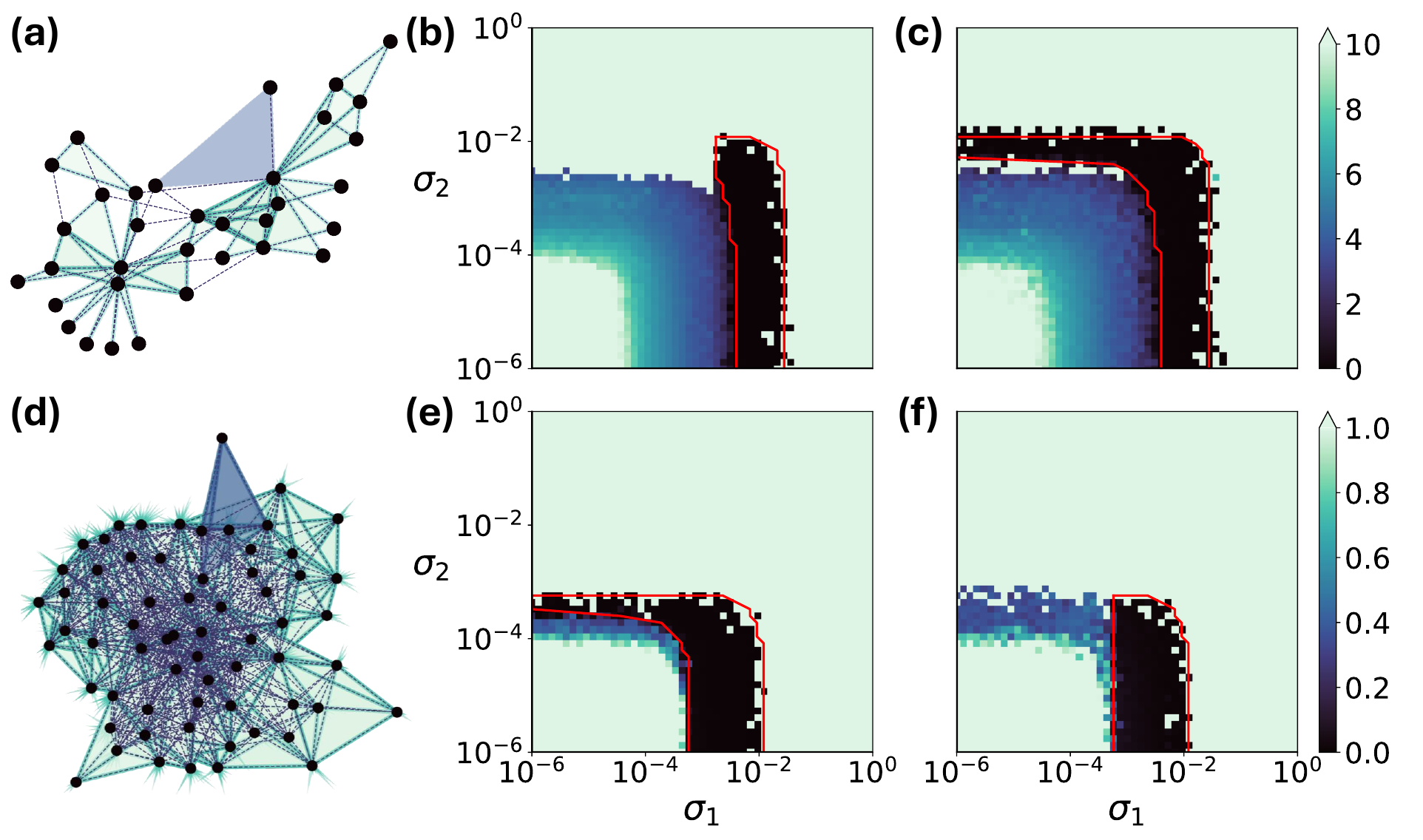}
\vspace{1cm}\caption{\textbf{Connectedness of $2$-hypererdges shapes the region of synchronization stability in real-world structures.} (a)-(c) Hypergraph from the Zachary Karate Club dataset, with an additional $2-$hyperedege shown in blue. (a) Graphical representation of the structure. (b) Region of synchronization stability (black area) for the original hypergraph. (c) Region of synchronization stability for the hypergraph with the further $2-$hyperedge added to reach full connectedness at the level of $2-$hyperedges. (d)-(f) Hypergraph from the cat brain's connectome. (d) Graphical representation of the structure. (e) Region of synchronization stability for the original hypergraph. (f) Region of synchronization stability for the hypergraph after isolation of a node, removing the $2-$hyperedges to which it belongs. In all panels, a system of Rössler oscillators with $\textbf{h}^{(1)}(\textbf{x}_j)=[x_j^3,0,0]^T$ and $\textbf{h}^{(2)}(\textbf{x}_j,\textbf{x}_k)=[x_j^2x_k,0,0]^T$ is considered. The red curves represent the theoretical prediction by the master stability function approach, while the color code corresponds to the outcome of the numerical simulations. Note that, in panel (c), the soft green points below the black area result from numerical instabilities in the integration of the chaotic dynamics.}
\label{fig:P8}
\end{figure*}

\subsection{The qualitative impact of intra-order hyperedge overlap}
\label{subsec:IVC}


So far, in both Secs. \ref{subsec:IVA} and \ref{subsec:IVB} we have fixed the coupling strengths $\sigma^{(1)}$ and $\sigma^{(2)}$. However, it is also important to analyse the effect of the hyperedge overlap within a continuous variation of these parameters. We carry out this analysis for the two configurations with extreme values of the overlap, namely $\left\{\mathcal{T}^{(2)}=0,\mathcal{I}^{(1,2)}=1\right\}$ and $\left\{\mathcal{T}^{(2)}=1,\mathcal{I}^{(1,2)}=1\right\}$. The results are shown in Fig.~\ref{fig:P7}(a)-(b), which illustrates the stability region predicted by the master stability function for Eq.~(\ref{eq:caseIIIsystem}) in the parameter space ($\sigma^{(1)}-\sigma^{(2)}$).
Remarkably, the two configurations show different shapes for the stability region, indicating a qualitatively distinct behavior associated to the two systems. For the structure with minimum intra-order hyperedge overlap, the synchronous state can be stable even when $\sigma_1$ is exceedingly small or zero, indicating the presence of weak (or even null) pairwise interactions. 
We observe that synchronization can be induced though a fine tuning of the coupling strength associated to three-body interactions, $\sigma_2$, since there exist lower and upper bounds of this parameter that guarantee synchronization stability. In contrast, for the structure with maximal intra-order hyperedge overlap, the first boundary (marking the transition from incoherent behavior to synchronization) becomes almost vertical such that for low values of $\sigma_1$ synchronization becomes impossible to achieve. The qualitative difference of the two scenarios can be exemplified by fixing a small value for $\sigma_1$, e.g., $\sigma_1=10^{-4}$, and letting $\sigma_2$ increases from zero. This corresponds to move, in the diagrams of Fig.~\ref{fig:P7}(a)-(b), from the bottom to the upper part, along the vertical line $\sigma_1=10^{-4}$. When the structure has no intra-order correlations as in Fig.~\ref{fig:P7}(a), there is a finite region of stability, consistently with the typical behavior of class III systems, whereas, when the degree of overlap in the structure is larger as in Fig.~\ref{fig:P7}(b), synchronization is impossible to obtain regardless of the value of the coupling strength $\sigma_2$. The latter behavior resembles that associated with class I systems on pairwise networks, for which the system dynamics remains incoherent irrespectively of the strength of the interactions. 

To shed more light on this important qualitative difference in the dynamical behavior, let $\Omega$ be the area of the region of stability, namely the black region of Fig.~\ref{fig:P7}(a) or (b). 
More precisely, the quantity $\Omega$ is computed by integrating the contour function of the stability region, denoted as $c(x,y)$, after a change of reference frame, from logarithmic scale $\sigma^{(1)},\sigma^{(2)}\in[10^{-6},10^0]$ ($m=1,2$) to linear scale $x,y\in[0,1]$ :
\begin{equation}
\Omega=\int_0^1\int_0^1c(x,y)dxdy
\label{eq:Omega}
\end{equation}

The average value of $\Omega$, computed by considering 100 hypergraphs for each pair of values of the intra-order and intra-order hyperedge ovelap, is shown in Fig.~\ref{fig:P7}(c). 
We notice that, for high values of the intra-order hyperedge overlap, the contour lines for $\Omega=0.25$ and $\Omega=0.27$ look similar to the ones of the eigenvalue ratio shown in Fig.~\ref{fig:P6}(c). Conversely, for low values of intra-order hyperedge overlap, the contour lines are nearly vertical, pinpointing that inter-order hyperedge overlap has a small impact. The change in the behaviour seems to occur at a specific value of $\mathcal{T}^{(2)}_c$, which only depends on the $2$-hyperedges. These considerations are confirmed in Fig.~\ref{fig:P7}(d), which shows the algebraic connectivity of the generalized Laplacian matrices restricted to $2$-hyperedges, namely $\lambda_2(L^{(2)})$, for each of the sets of 100 structures considered for each pair of values of the two hyperedge overlaps. For structures having $\mathcal{T}^{(2)}>\mathcal{T}^{(2)}_c$, we find that $\lambda_2(L^{(2)})=0$. As for Laplacian matrices, the multiplicity of the zero eigenvalue corresponds to the number of connected components, we conclude that, in this case, the structure obtained using only the existing $2$-hyperedges is not connected. Since setting very small values of $\sigma^{(1)}$ (in the limit case, $\sigma^{(1)}\rightarrow 0$), means to rely exclusively on the interactions associated with the set of $2-$hyperedges, stability of the synchronous state is impossible to achieve. 

On the contrary, when $\mathcal{T}^{(2)}<\mathcal{T}^{(2)}_c$, the values of the algebraic connectivity exhibit a bimodal distribution between the values $\lambda_2(L^{(2)})=0$ (obtained in disconnected sets of $2$-hyperedges) and the values $\lambda_2(L^{(2)})\neq0$ (obtained in connected sets of $2$-hyperedges). This bimodality is captured by the parameter $\phi$ (green-filled curve in Fig.~\ref{fig:P7}(d)), which accounts for the fraction of structures with a given $\mathcal{T}^{(2)}$ and $\lambda_2(L^{(2)})=0$, and by the curve shown in Fig.~\ref{fig:P7}(e). This latter panel shows $\Omega$ as a function of the intra-order and inter-order hyperedge overlap, clearly demonstrating the bimodality of the distribution of $\Omega$ for $\mathcal{T}^{(2)}<\mathcal{T}^{(2)}_c$. Within each of the two modes, low (high) values of overlap enlarge (reduce) the synchronization area. This is consistent what observed in Fig.~\ref{fig:P6}(c). Furthermore, we note that, although the bimodality (as well as the value of $\mathcal{T}^{(2)}_c$) depends on the rewiring procedure used to create the synthetic network under analysis, in any case, to obtain the maximum intra-order overlap for an order $m$, the set of $m-$hyperedges must be disconnected. Conversely, the random distribution of $m$-hyperedges is unlikely to result in a disconnected set.


Lastly, we show that the two different behaviors discussed above also appear in real-world structures. For the purpose of exemplification, we use hypergraphs obtained from two real-world datasets, namely the Zachary Karate Club~\cite{zachary1977information} and a cat's brain connectome~\cite{de2013rich}. For both structures, we consider three generic nodes $i$, $j$, and $k$ to be connected by an $2-$hyperedge, each time we find in the original dataset a clique involving them~\cite{gambuzza2021stability,zhang2023higher}. This yields a structure having $\mathcal{I}^{(1,2)}=1$.

Fig.~\ref{fig:P8}(a) shows the hypergraph obtained starting from the Zachary Karate Club dataset, with the green triangles representing $2-$hyperedges. The synchronization region predicted by the master stability function for this hypergraph is depicted in Fig.~\ref{fig:P8}(b). Here, we observe that synchronization cannot be obtained when the coupling strength associated to pairwise interactions is small (e.g., $\sigma_1\rightarrow0$). This is due to the fact that two nodes, highlighted by blue circles in Fig.~\ref{fig:P8}(a), do not belong to any $2-$hyperedge. The introduction of an additional $2-$hyperedge (depicted in Fig.~\ref{fig:P8}(a) as a blue triangle) in the structure makes it connected also at the level of $2-$hyperedges, resulting in a bounded region of synchronization stability also when $\sigma_1\rightarrow0$ (Fig.~\ref{fig:P8}(c)). 

For the hypergraph obtained from the cat connectome dataset (Fig.~\ref{fig:P8}(d)) we find a different scenario since all nodes are connected through $2-$hyperedges. Consequently, the system has a finite stability region for small $\sigma_1$ (Fig.~\ref{fig:P8}(e)). Furthermore, upon isolating a node (highlighted with a blue circle) by removing the $2-$hyperedges to which it belongs (blue triangles in Fig.~\ref{fig:P8}(d)), the set of $2-$hyperedges is no longer connected. Consequently, a change of a single $2-$hyperedge results into a loss of stability for the synchronous state when $\sigma_1\rightarrow0$. 

\section{Discussion and conclusions}
\label{sec:V}

In this work, we have investigated the relationship between the microscopic organization of higher-order structures and the collective behavior in systems of coupled dynamical systems. To this end, we have introduced a general framework to characterize the hyperedge overlap of higher-order structures: the overlap matrix. This matrix has two types of elements: the diagonal matrix elements, quantifying the hyperedge overlap between interactions of the same order, and the non-diagonal elements, quantifying the hyperedge overlap between interactions of different orders. This framework offers a comprehensive tool to analyze all types and orders of hyperedge overlap within a structure, unlike other metrics that focus on a single measure combining all previous aspects (and thus not allowing to distinguish among them) \cite{lee2021hyperedges}, or solely on intra-order \cite{malizia2023hyperedge} or inter-order correlations \cite{landry2024simpliciality}.

Throughout the paper, we have examined the impact of hyperedge overlap on the stability of synchronization in systems of coupled dynamical systems. The stability is determined by the effective Laplacian, a matrix encapsulating the structure and strength of interactions across all orders. Our findings indicate that a high degree of hyperedge overlap hampers the stability of synchronization. Hence, since a high degree of overlap (in particular, inter-order) is associated with hyperedges satisfying the downward closure, our result suggest that random hyperedges promote stability more effectively than simplicial complexes. Furthermore, our findings pinpoint that large hyperedege overlap within an order of interaction promotes the tendency towards local synchronization. Moreover, a hierarchy among the elements of the overlap matrix emerges, where higher-order overlaps are more critical for synchronization stability. This occurs in two ways: first, larger values of intra-order hyperedge overlap result in a clustered organization at the mesoscale; second, in the case of non-zero inter-order hyperedge overlap, larger values of intra-order hyperedge overlap result in the lower orders of interactions being overlapped to some extent. 


By analyzing the behavior of coupled chaotic oscillators, we have found that intra- and inter-order hyperedge overlaps have a qualitatively distinct impact on the emerging dynamics. This difference originates from the fact that, for a given order of interactions, the set of hyperedges of that order must be disconnected in order to reach the maximum value of intra-order hyperedge overlap. Conversely, when the intra-order hyperedge overlap is at its minimum, the probability that the structure is disconnected is negligible. The lack of connectedness at the level of interactions of a given order interactions makes impossible to fulfill the constraint associated to synchronization stability when the interaction strength associated to interactions of this order dominates over the the ones associated to the other interactions. These findings, which we have first obtained on synthetic structures where all parameters were tunable, have been then exemplified and confirmed also in hypergraphs constructed from real systems.

Overall, our results underscore the significance of microscopic organization for the dynamics of systems with higher-order interactions, with particular reference to the analysis to dynamical systems of coupled oscillators, elucidating the independent and combined effects of both types of overlap. Future work could extend the analysis to more sophisticated systems such as pacemakers \cite{djabella2007two}, coupled neurons  \cite{izhikevich2003simple,izhikevich2007dynamical}, or higher-order nonlinear consensus dynamics \cite{neuhauser2020multibody,neuhauser2021consensus}.

\bigskip

\noindent\textbf{Data availability:} The brain connectome dataset was downloaded from https://neurodata.io/project/connectomes/ and the Karate club dataset \cite{zachary1977information} was downloaded through the \textit{networkx} library in Python.

\noindent\textbf{Code availability:} The code is available at https://github.com/santiagolaot/hyperedge-overlap.

\section*{Acknowledgements} 
The authors would like to thank Lucia Valentina Gambuzza for the data on the master stability function. S.L.O and J.G.G. acknowledge support from Departamento de Industria e Innovaweci\'on del Gobierno de Arag\'on y Fondo Social Europeo (FENOL group grant E36-23R) and Ministerio de Ciencia e Innovaci\'on (grant PID2020-113582GB-I00 and PID2023-147734NB-I00). S.L.O. acknowledges financial support from Gobierno de Aragón through a doctoral fellowship. V.L. acknowledges support from the European Union -  NextGenerationEU, in the framework of the Growing Resilient, INclusive and Sustainable project (GRINS PE00000018 – CUP E63C22002120006). 

\bigskip

\begin{figure}[t!]
\centering\includegraphics[width=0.85\linewidth]{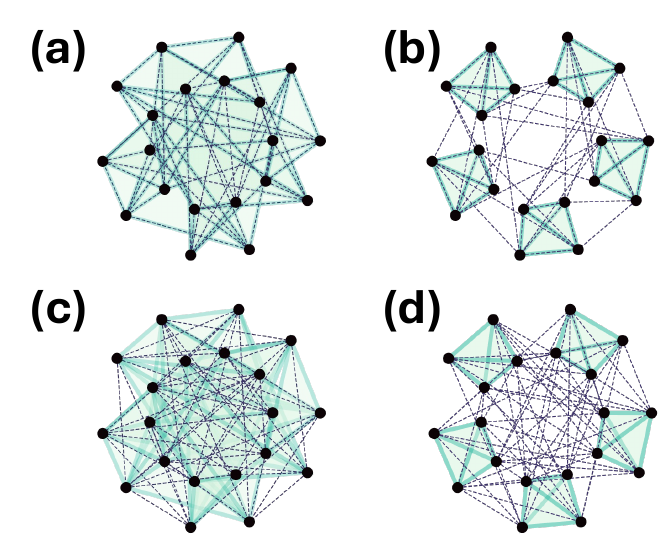}
\vspace{1cm}\caption{\textbf{Examples of synthetic structures with tunable hyperedge overlap.} Examples of structures with $N=20$ nodes and connectivity $k^{(1)}=6$ and $k^{(2)}=3$ in four extreme scenarios: (a) $\mathcal{I}^{(1,2)}=1$ and $\mathcal{T}^{(2)}=0$; (b) $\mathcal{I}^{(1,2)}=1$ and $\mathcal{T}^{(2)}=1$; (c) $\mathcal{I}^{(1,2)}=0$ and $\mathcal{T}^{(2)}=0$; (d) $\mathcal{I}^{(1,2)}=0$ and $\mathcal{T}^{(2)}=1$.
}
\label{fig:PApp}
\end{figure}

\section*{APPENDIX A: SYNTHETIC STRUCTURES}
\label{sec:appendixA}

In order to obtain sets of hypernetworks covering the $\{\mathcal{T}^{(2)},\mathcal{I}^{(1,2)}\}$-space, we first generate a set of 2-hyperedges with $k_i^{(2)} = k^{(2)}$ for $i = 1, \ldots, N$, ensuring maximum intra-order hyperedge overlap ($\mathcal{T}^{(2)}=1$) using the procedure described in \cite{malizia2023hyperedge}.
Next, we apply a numerical rewiring to the set of 2-hyperedges to minimize the intra-order hyperedge overlap while maintaining the same initial generalized degree, thus obtaining sets of 2-hyperedges spanning the full range $\mathcal{T}^{(2)} \in [0,1]$. For each set of 2-hyperedges created, we consider a set of 1-hyperedges (links), where all nodes share the same number of neighbors $k^{(1)}$, that cover all the faces of the 2-hyperedges layer. This will result in hypernetworks with maximized inter-order hyperedge overlap values for any given scenario of $\mathcal{T}^{(2)}$.
Subsequently, we rewire the 1-hyperedges to minimize the inter-order hyperedge overlap for each hypernetwork, creating an ensemble that covers the entire parameter space of $\{\mathcal{T}^{(2)}, \mathcal{I}^{(1,2)}\}$.  In Fig.~\ref{fig:PApp}(a)-(d), we illustrate four graphical examples of hypernetworks with different configurations of both metrics. 


%
%
%

\bibliography{biblio}

\end{document}